\documentclass[aps, showpacs, showkeys,nofootinbib,floatfix]{revtex4}

\usepackage{amssymb}
\usepackage{amsmath}
\usepackage{graphicx}

\begin{document}

\title{CMB Temperature and Matter Power Spectrum in a Decay Vacuum Dark Energy Model}

\author{Lixin Xu$^{1,2}$}
\email{lxxu@dlut.edu.cn}
\author{Yuting Wang$^2$}
\author{Minglei Tong$^1$}
\author{Hyerim Noh$^1$}

\affiliation{$^1$Korea Astronomy and Space Science Institute,
Yuseong Daedeokdaero 776,
Daejeon 305-348,
R. Korea}
\affiliation{$^2$Institute of Theoretical Physics, School of Physics \&
Optoelectronic Technology, Dalian University of Technology, Dalian,
116024, P. R. China}

\begin{abstract}
In this paper, a decay vacuum model $\bar{\rho}_\Lambda=3\sigma M_p^2H_0 H$ is revisited by detailed analysis
of background evolution and perturbation equations. We show the
imprints on CMB temperature and matter power spectrum from the effective coupling terms between dark sectors by comparing to
the standard cosmological constant model and observational data points (WMAP7 and SDSS DR7). We find that the decay vacuum model can describe the expansion rate at late times as well
as the standard cosmological constant model but it fails to simultaneously
reproduce the observed CMB and matter power spectrum. Its generalization $\bar{\rho}_\Lambda=3M_p^2(\xi_1 H_0 H+\xi_2  H^2)$ is also discussed. Detailed analysis of the background evolution shows that the dimensionless parameter $\xi_{2}$ would be zero to avoid the unnatural 'fine tuning' and to keep the positivity of energy density of dark matter and dark energy in the early epoch.
\end{abstract}

\pacs{Added}

 \hfill XU-KASI/01

\maketitle

\section{Introduction}
Since the accelerating expansion of the universe has been found from the measures of the luminosity-redshift relation $d_L(z)$ of type Ia supernovae (SN Ia) \cite{Riess}, a cosmic component called dark energy was often introduced to explain the acceleration within the framework of general relativity. Now more and more evidence, such as cosmic microwave background (CMB) \cite{Bennett,ref:wmap7}, baryon acoustic oscillations (BAO) \cite{Eisenstein}, weak gravitational lensing \cite{Jarvis} and x-ray clusters \cite{Allen,Allen2}, indicated that  the universe is spatially flat and dominated by dark energy at present. Apart from dark energy models, modified gravity \cite{Carroll,Trodden} can also explain the acceleration of the present universe. However, we just focus on dark energy model in this paper. Among the various dark energy models \cite{Copeland}, including scalar field \cite{Ratra1}, vector field\cite{Zhang94,Kiselev}, holographic dark energy \cite{Holographic}, Chaplygin gas \cite{Kamenshchik} and so on. The cosmological constant model ($\Lambda$CDM) \cite{Weinberg} is the simplest one. However, as is well known, the $\Lambda$CDM suffers from a fine tuning problem: the observed vacuum energy density of order $\sim10^{-47}\text{GeV}^4$ is about $10^{121}$ orders of magnitude smaller than the value expected by quantum field theory for a cutoff scale that is the Plank scale, and is still about $10^{44}$ orders smaller even for a cutoff scale that is the QCD scale \cite{Copeland}. As an extension to $\Lambda$CDM, the decaying vacuum (DV) dark energy model was  proposed \cite{Borges1,Carneiro}, based on the incomplete quantum field theory in the curved 4-dimension space-time. In this model, the vacuum serves as  dark energy, whose energy density decays with the expansion of the universe leading to an additional production of the matter component. In the late-time with a quasi-de Sitter background, the vacuum density is proportional to the Hubble rate, $\rho_\Lambda(t) \propto H(t)$. However, the equation of state for the vacuum is a constant value $w=p_\Lambda(t)/\rho_\Lambda(t)=-1$, the same as that in the $\Lambda$CDM model. Moreover, as an interesting feature, the late-time dynamics of the DV model is similar to $\Lambda$CDM \cite{Borges1,Carneiro}.

The quasar APM 08279+5255 at $z=3.91$ was used to examine the DV model \cite{mltong2}, and it was found that the DV model can greatly alleviate the high redshift cosmic age problem existing in the $\Lambda$CDM model. In order to distinguish the DV model from other dark energy models at the late-time Universe, the state finder and $Om$ diagnostics of the DV model were also presented in \cite{mltong2}. Moreover, the DV model has been tested by $\chi^2$ analysis using the observational data of SN Ia \cite{Carneiro2}, a joint data from SN Ia, BAO and CMB\cite{Carneiro3,Pigozzo}, and the joint data that the Gamma-ray bursts, Hubble rate and x rays in galaxy clusters were added \cite{tongnoh}. It was found that, the DV model favors a relatively larger value of the matter density contrast, $\Omega_m=(0.34\sim0.43)$.
 
In this paper, compared to previous work \cite{Pigozzo,tongnoh}, we take the radiation component into account in a more reasonable way. We will demonstrate the temperature anisotropies of CMB induced by the matter perturbations in the DV model with various values of $\Omega_m$. The matter power spectrum is investigated.
 
This paper is structured as follows. In section \ref{sec:bg}, the background evolution equations are given in a spatially flat FRW universe. We give a brief review of cosmological perturbation theory in section \ref{sec:cp}. The main results of this paper are presented in section \ref{sec:results}. In section \ref{sec:tvv}, we give a brief discussion on a generalized form  $\bar{\rho}_{\Lambda}=3M_p^2(\xi_1 H_0 H+\xi_2  H^2)$ and point out that this model is not viable. Section \ref{sec:con} is the conclusion.

\section{Background Evolution Equations}\label{sec:bg}

\begin{figure}[!htbp]
\includegraphics[width=5.3cm]{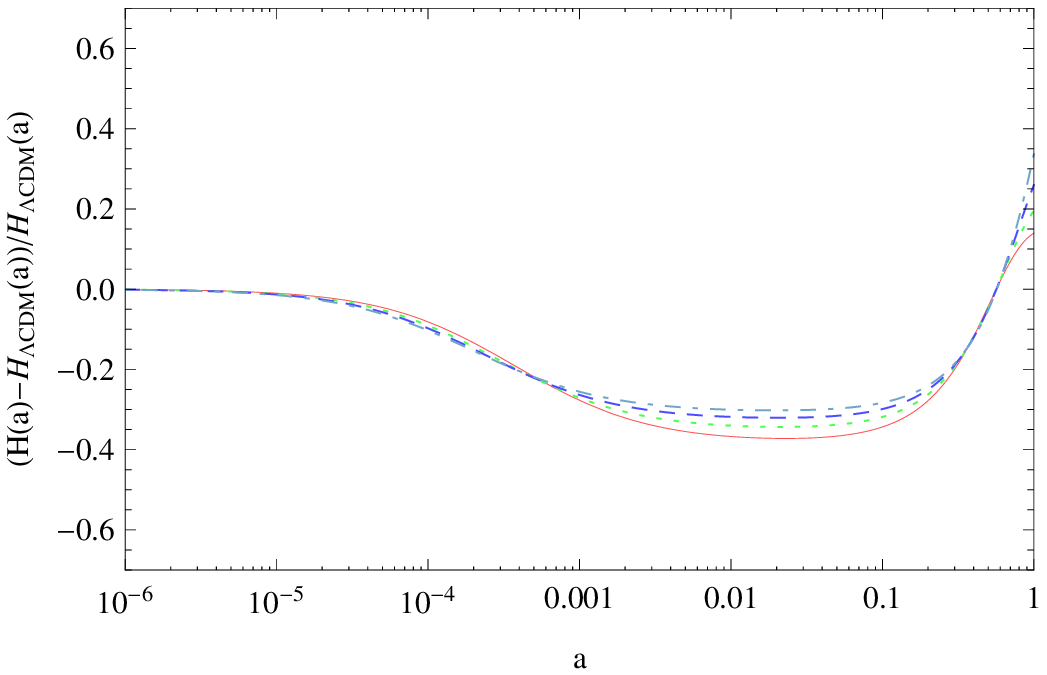}
\includegraphics[width=5.2cm]{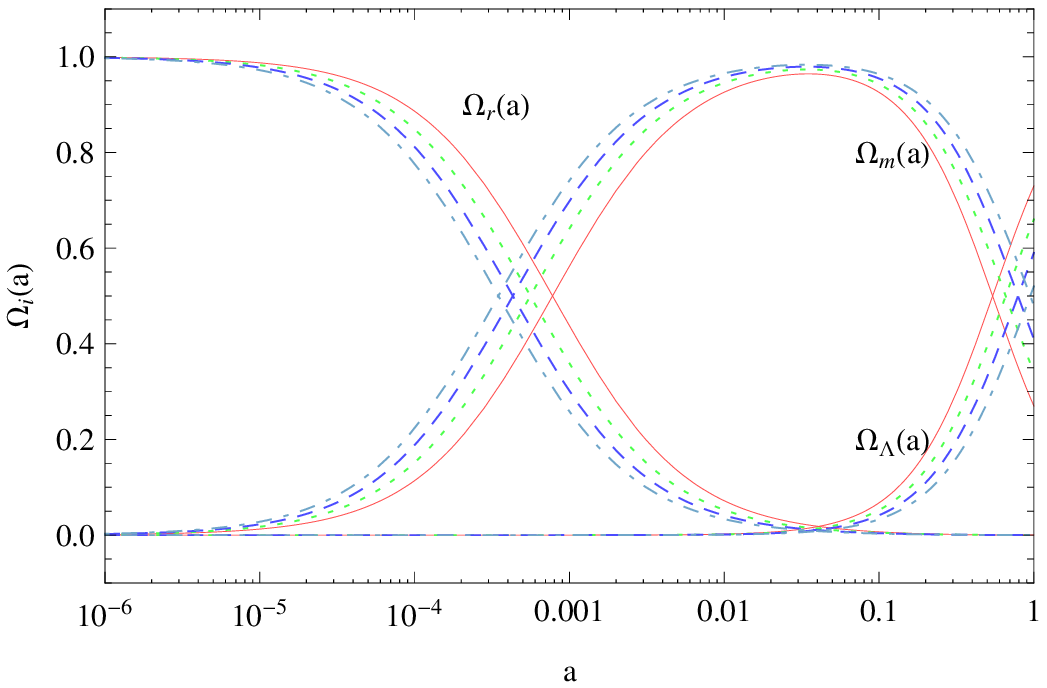}
\includegraphics[width=5.2cm]{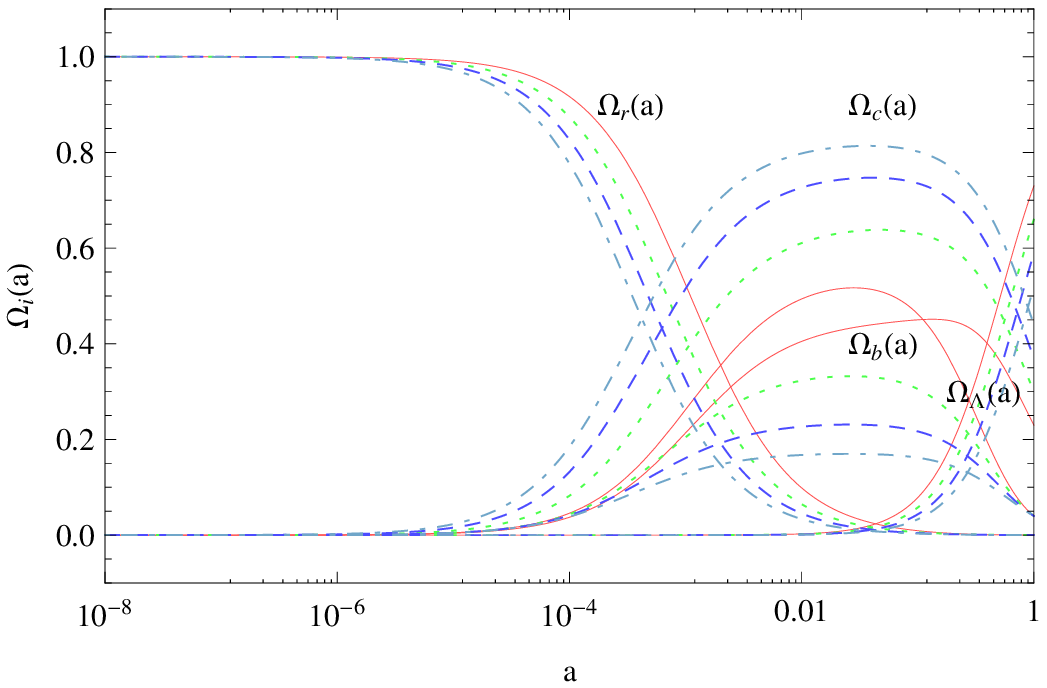}
\caption{Left panel: The relative deviations of Hubble parameter $H(a)$ to $\Lambda$CDM model, i.e. $(H(a)-H_{\Lambda CDM}(a))/H_{\Lambda CDM}(a)$ with respect to scale factor $a$. Central and Right panels: the evolutions of dimensionless energy components with respect to scale factor $a$. The solid (red),  dotted (green), dashed (blue) and dash-dotted (dark green) lines correspond to the values of $\Omega_{c0}=0.23, 0.30, 0.37, 0.44$ respectively. Here $H_0=72$, $\Omega_{b0}=0.04$ and $\Omega_{r0}=0.00008$ are adopted.}\label{fig:evolution}
\end{figure}

The energy density of the decay vacuum is given as \cite{Borges1} 
\begin{equation}
\bar{\rho}_{\Lambda}=3\sigma M_p^2H_0 H. 
\end{equation}
 Here $M_p^{-2}=8\pi G$ is the reduced Plank mass. An extra $H_0$ is introduced for convenience. And $\sigma$ is a positive dimensionless parameter. The energy-momentum conservation equation implies the interaction between dark matter and decay vacuum,
\begin{eqnarray}
\dot{\bar{\rho}}_c+3 H\bar{\rho}_c=-\bar{Q},\\
\dot{\bar{\rho}}_\Lambda+3
H(\bar{\rho}_\Lambda+\bar{p}_\Lambda)=\bar{Q}.
\end{eqnarray}
where $\bar{Q}$ denotes an interaction bewteen dark sectors. In this paper, a non-gravitational interaction between dark sectors is considered only. The other case where the non-gravitational interaction between dark sectors and radiation is considered by other authors, for example \cite{ref:DVMR}. The remaining energy components $\bar{\rho}_{b}$ and $\bar{\rho}_{r}$ respect the usual conservation equation $\dot{\bar{\rho}}_{i}+3H(1+w_i)\bar{\rho}_{i}=0$, where $i=b,r$ and $w_{b}=0$, $w_{r}=1/3$ are the equation of state of baryon and radiation respectively. The Friedmann equation in a spatially flat FRW universe is given as
\begin{equation}
H^2=\frac{1}{3M^2_p}\left(\bar{\rho}_r+\bar{\rho}_b+\bar{\rho}_c+\bar{\rho}_\Lambda\right),
\end{equation}
where the subscripts $r, b, c,\Lambda$ denote radiation, baryon, cold dark matter and vacuum energy density respectively. One can rewrite the Friedmann equation into
\begin{equation}
H^2=H^2_0\left[\Omega_{r0}a^{-4}+\Omega_{b0}a^{-3}+\Omega_{c0}f_c(a)+(1-\Omega_{r0}-\Omega_{b0}-\Omega_{c0})f_{de}(a)\right],\label{eq:FRE}
\end{equation}
where $\Omega_{i}=\frac{\rho_i}{3M^2_pH^2}$ is the dimensionless energy parameter of $i=r,b,c,\Lambda$ component, $f_c(a)=\bar{\rho}_c/\bar{\rho}_{c0}$ and $f_{de}(a)=\bar{\rho}_{de}/\bar{\rho}_{de0}$
are fractions of dark matter and dark energy respectively. Hereafter, the subsript '$0$' denotes the corresponding value at present, i.e., the corresponding value at scale factor $a=1$. And the terms $\Lambda$ and dark energy are exchangeable. Clearly, one has the present values of these fractions $f_c(1)\equiv1$ and $f_{de}(1)\equiv1$ respectively. For a spatially flat universe, one has $\Omega_{de0}=\sigma=1-\Omega_{r0}-\Omega_{b0}-\Omega_{c0}$, i.e., the relation
\begin{equation}
\sigma=1-\Omega_{r0}-\Omega_{b0}-\Omega_{c0}.
\end{equation}
Considering the interaction between dark sectors, one
has the evolution equations of $f_{c}(a)$ and $f_{de}(a)$
\begin{eqnarray}
\frac{d f_{c}}{d\ln a}+3 f_c=-\frac{\bar{Q}}{3M^2_p H_0^2H\Omega_{c0}},\\
\frac{d f_{de}}{d\ln a}+3
(1+w_{de})f_{de}=\frac{\bar{Q}}{3M^2_pH_0^2H\Omega_{de0}},
\end{eqnarray}
where $w_{de}=\bar{p}_{de}/\bar{\rho}_{de}$ is the equation of state (EoS) of dark energy. For cosmological constant, $w_{de}\equiv-1$. It means that the interaction term is
\begin{equation}
\bar{Q}=\dot{\bar{\rho}}_{\Lambda}=3\sigma M^2_p H_0\dot{H}=3\sigma M^2_p a H_0 H\frac{d H}{d a},
\end{equation}
and the fraction of dark energy is
\begin{equation}
f_{de}=\frac{\bar{\rho}_{de}}{\bar{\rho}_{de0}}\equiv H/H_0.
\end{equation}
Substituting the above relation of $f_{de}$ into Eq. (\ref{eq:FRE}), one has a quadratic equation of $H/H_0$. After a simple algebra, one obtains
\begin{equation}
H=\frac{1}{2}H_0\left[\Omega_{de0}+\sqrt{\Omega^2_{de0}+4(\Omega_{r0}a^{-4}+\Omega_{b0}a^{-3}+\Omega_{c0}f_c(a))}\right]\label{eq:H}.
\end{equation}
Here, the other solution is removed for its negativity. The function $f_c(a)$ is a solution of the differential equation
\begin{equation}
\frac{d f_c(a)}{d\ln a}=\frac{\Omega_{de0}(3\Omega_{b0}a^{-3}+4\Omega_{r0}a^{-4})-3\Omega_{c0}f_c(a)\sqrt{\Omega_{de0}^2+4(\Omega_{r0}a^{-4}+\Omega_{b0}a^{-3}+\Omega_{c0}f_c(a))}}{\Omega_{c0}\left[\Omega_{de0}+\sqrt{\Omega_{de0}^2+4(\Omega_{r0}a^{-4}+\Omega_{b0}a^{-3}+\Omega_{c0}f_c(a)})\right]}
\end{equation}
with current value $f_c(1)\equiv1$. It is easy to obtain the conventional dark matter evolution equation $f_c(a)=a^{-3}$, when the cosmological constant is a real constant. From the Friedmann equation, one has
\begin{equation}
\frac{d H}{d a}=-\frac{H_0\left(3\Omega_{b0}a^{-3}+3\Omega_{c0}f_c(a)+4\Omega_{r0}a^{-4}\right)}{a\left(\Omega_{de0}+\sqrt{\Omega_{de0}^2+4(\Omega_{r0}a^{-4}+\Omega_{b0}a^{-3}+\Omega_{c0}f_c(a)})\right)}.
\end{equation}
When one omits the radiation and baryon components the Friedmann equation
\begin{equation}
H=H_0(1-\Omega_{c0}+\Omega_{c0}a^{-3/2})
\end{equation}
is recovered \cite{mltong2}. In the previous study, when one considers the early stage of universe, the approximated Friedmann equation
\begin{equation}
H(a)\approx H_0\left[\left(1-\Omega_{m0}+\Omega_{m0}a^{-3/2}\right)^2+\Omega_{r0}a^{-4}\right]^{1/2}
\end{equation}
is adopted \cite{mltong2}. As a contrast, in this paper, an exact one is obtained. We plot the evolution curves for illustrating the relative deviation from $\Lambda$CDM model with different values of $\Omega_{c0}$ in the left panel of Fig. \ref{fig:evolution}. The corresponding right panel shows the evolution of dimensionless energy parameters $\Omega_{i}, i=b, c, r, \Lambda$ with respect to scale factor $a$.
It shows the relative deviation from $\Lambda$CDM model. One can read that the Hubble parameter is less than that of $\Lambda$CDM model in almost the whole history of the universe, with the exception of recent epoch.  In this case, the geometric probes, for example the luminosity distance, would not distinguish it from $\Lambda$CDM model due to the negative and positive mixture of relative deviations. The central panel shows the evolutions of dimensionless parameters $\Omega_{r}$, $\Omega_{m}=\Omega_{c}+\Omega_{b}$ and $\Omega_{\Lambda}$. From the central panel, when the cold dark matter and baryon are combined as a whole matter component $\Omega_{m}$, one may see that the components evolve as usual naively. However, from the right panel, one sees that the evolution of $\Omega_{b}$ is almost comparable to $\Omega_{c}$ when $\Omega_{c0}\sim 0.23$. One can also read the difference from the left panel of Fig \ref{fig:evolution}. The right panel implies the lower abundance of $\Omega_{b}$ and high $\Omega_{c}$ is favored in this decay vacuum model. But, the higher abundance of $\Omega_{b}$ and lower one of $\Omega_{c}$ will make the dynamic evolution different. To distinguish this decay vacuum model from $\Lambda$CDM model, one should
consider high redshift observational data points and dynamical
evolution. The observation of CMB is a good indicator for its high
redshift ($z\sim 1089$) and precision ($\Delta T/T_{CMB}\sim10^{-5}$). So, in this paper, we are going to
investigate the observational effects under this decay vacuum model. On the other aspect, if the geometric evolution history is very similar, but the dynamic evolution would be very different. In this decay vacuum case, an effective interaction is introduced for keeping the conservation of energy and momentum. In this sense, the matter power spectrum is also investigated.

\section{Cosmological Perturbation Equations}\label{sec:cp}

In this section, at first, we give a brief review of cosmological perturbation theory with an arbitrary interaction between dark sectors. For the cosmological perturbation theory, please see \cite{ref:CP} and references therein. For the gauge ready formalism about the perturbation theory, please see \cite{ref:Hwang}. Here, we follow the work \cite{ref:Viliviita}  closely. For considering decay vacuum, we assume the dark energy is a smooth energy component. It means that the dark matter perturbation equation is modified due to an interaction between dark sectors.  

Scalar perturbations of the flat FRW metric are given in the following form \cite{ref:Viliviita} 
\begin{equation}
ds^2=a^2\left\{-(1+2\phi)d\tau^2+2\partial _i B d\tau dx^i
+\left[(1-2\psi)\delta_{ij}+2\partial_i\partial_j E\right]dx^i
dx^j\right\}.
\end{equation}
$\bar{u}^\mu=a^{-1}(1,0,0,0)$ is the background four-velocity. Its spatial part is the perturbation, we can set it as $\partial^i v_A$ for the corresponding scalar perturbation only. Then using the equality $g_{\mu\nu}u_A^\mu u_A^\nu=-1$, one has the four-velocity of A-fluid \cite{ref:Viliviita} 
\begin{equation}
u^\mu_{A}=a^{-1}(1-\phi, \partial^i v_A),\quad u^A_\mu=g_{\mu\nu}u^\nu_{A}=a(-1-\phi,\partial_i [v_A+B]),
\end{equation}
where $v_A$ is the peculiar velocity potential. The local volume expansion rate is $\theta = \overrightarrow{\nabla}\cdot \overrightarrow{v}$ which is $\theta_A=-k^2(v_A+B)$ in Fourier space.

The perturbed energy-momentum tensor is given as
\begin{eqnarray}
\delta\nabla_{\mu} T^{\mu 0}_A=\frac{1}{a^2}\left\{\delta \rho'_A+3\mathcal{H}(\delta\rho_A+\delta p_A)-3(\bar{\rho}_A+\bar{p}_A)\psi'+(\bar{\rho}_A+\bar{p}_A)\nabla^2(v_A+E')\right.\nonumber\\
\left.-2\left[\bar{\rho}'_A+3\mathcal{H}(\bar{\rho}_A+\bar{p}_A)\right]\phi\right\},\\
\delta \nabla_{\mu}T^{\mu i}_A=\frac{1}{a^2}\partial^i\left\{\left[(\bar{\rho}_A+\bar{p}_A)(v_A+B)\right]'+4\mathcal{H}(\bar{\rho}_A+\bar{p}_A)(v_A+B)+(\bar{\rho}_A+\bar{p}_A)\phi+\delta p_A\right.\nonumber\\
\left.+\frac{2}{3}\bar{p}_A\nabla^2\pi_A
-\left[\bar{\rho}'_A+3\mathcal{H}(\bar{\rho}_A+\bar{p}_A)\right]B\right\}.
\end{eqnarray}
where prime '$'$' denotes derivative with respect to the conformal time $d\eta=dt/a(t)$, and $\mathcal{H}=a'/a$ is the conformal Hubble parameter.
When the interaction between the fluids is introduced, the energy-momentum conservation equation becomes \cite{ref:Viliviita} 
\begin{equation}
\nabla_\mu T^{\mu\nu}_A=Q_A^{\nu},\quad \delta\nabla_\mu
T^{\mu\nu}_A=\delta Q_A^{\nu}.
\end{equation}
The background evolution equation of fluid-A is
\begin{equation}
\bar{\rho}'_A+3\mathcal{H}(\bar{\rho}_A+\bar{p}_A)=a\bar{Q}_A,
\end{equation}
where $\bar{Q}_A$ is the background term of the general interaction \cite{ref:Viliviita} 
\begin{equation}
Q^{\mu}_A=Q_A u^{\mu}+F^{\mu}_A,
\end{equation}
where
\begin{equation}
Q_A=\bar{Q}_A+\delta Q_A,\quad F^{\mu}_A=a^{-1}(0,\partial^i f_A).
\end{equation}
are energy and momentum transfer rates respectively. Then, one has the components
\begin{eqnarray}
Q^0_A=(\bar{Q}_A+\delta Q_A)u^0&=&a^{-1}(1-\phi)(\bar{Q}_A+\delta Q_A),\\
\delta Q^0_A&=&a^{-1}(\delta Q_A-\phi \bar{Q}_A),
\end{eqnarray}
and
\begin{eqnarray}
Q^i_A=(\bar{Q}_A+\delta Q_A)u^i+a^{-1}f_A
&=&a^{-1}(\bar{Q}_A+\delta Q_A)\partial^iv+a^{-1}\partial^i f_A,\\
\delta Q^i_A&=&a^{-1}\partial^i(\bar{Q}_A v+f_A).
\end{eqnarray}

Considering the interaction between the fluids, the perturbed energy and momentum balance equations are \cite{ref:Viliviita} 
\begin{eqnarray}
\delta \rho'_A&+&3\mathcal{H}(\delta\rho_A+\delta p_A)-3(\bar{\rho}_A+\bar{p}_A)\psi'+(\bar{\rho}_A+\bar{p}_A)\nabla^2(v_A+E')\nonumber\\
&=&a\bar{Q}_A\phi+a\delta Q_A,\\
\delta p_A&+&\left[(\bar{\rho}_A+\bar{p}_A)(v_A+B)\right]'+4\mathcal{H}(\bar{\rho}_A+\bar{p}_A)(v_A+B)+(\bar{\rho}_A+\bar{p}_A)\phi+\frac{2}{3}\bar{p}_A\nabla^2\pi_A\nonumber\\
&=&a\bar{Q}_A(B+v)+a f_A.
\end{eqnarray}

To solve the above equations or make them complete, one needs the relations between $\delta p_A$ and $\delta \rho_A$. The sound speed $c^{2}_{sA}$ of $A$ fluid is defined in the $A$ rest frame \cite{ref:Viliviita} 
\begin{equation}
c^2_{sA}=\frac{\delta p_A}{\delta \rho_A}|_{rf},
\end{equation}
where '$|_{rf}$' denotes the rest frame. The 'adiabatic sound speed' for any medium is defined as \cite{ref:Viliviita} 
\begin{equation}
c^2_{aA}=\frac{p'_A}{\rho'_A}=w_A+\frac{w'_A}{\rho'_A/\rho_A}.
\end{equation}
In the $A$ rest frame one has \cite{ref:Viliviita} 
\begin{equation}
T^i_0|_{rf}=0=T^0_i|_{rf}.
\end{equation}
To obtain the expression in a general gauge, one makes a gauge transformation, $x^\mu\rightarrow x^\mu+(\delta \tau_A, \partial^i\delta x_A)$ \cite{ref:Viliviita} , 
\begin{equation}
v_A+B=(v_A+B)|_{rf}+\delta \tau_A, \quad \delta p_A=\delta p_A|_{rf}-p'_A\delta \tau_A,\quad \delta \rho_A=\delta \rho_A|_{rf}-\rho'_A\delta \tau_A.
\end{equation}
Thus, one has $\tau_A=v_A+B$ and \cite{ref:Viliviita} 
\begin{eqnarray}
\delta p_A&=&\delta p_A|_{rf}-p'_A\delta \tau_A\nonumber\\
&=&c^2_{sA}\delta \rho_A+(c^2_{sA}-c^2_{aA})\rho'_A(v_A+B)\nonumber\\
&=&c^2_{aA}\delta \rho_A+\delta p_{nad A},
\end{eqnarray}
where $\delta p_{nad A}=(c^2_{sA}-c^2_{aA})\left[\delta \rho_A+\rho'_A(v_A+B)\right]$ is the intrinsic non-adiabatic perturbation in the $A$-fluid. When the interaction is introduced, the conservation equation becomes $\bar{\rho}'_A=-3\mathcal{H}(\bar{\rho}_A+\bar{p}_A)+a\bar{Q}_A$. By using the relation $\theta_A=-k^2(v_A+B)$ in Fourier space, one has \cite{ref:Viliviita} 
\begin{eqnarray}
\delta p_A&=&c^2_{sA}\delta \rho_A+(c^2_{sA}-c^2_{aA})\rho'_A(v_A+B)\nonumber\\
&=&c^2_{sA}\delta \rho_A+(c^2_{sA}-c^2_{aA})\left[3\mathcal{H}(1+w_A)\bar{\rho_A}-a\bar{Q}_A\right]\frac{\theta_A}{k^2}.
\end{eqnarray}

Defining the density contrast $\delta_A=\delta \rho_A/\bar{\rho}_A$, one has the evolution equations for density perturbation and velocity perturbations for a generic fluid \cite{ref:Viliviita} 
\begin{eqnarray}
\delta'_A&+&3\mathcal{H}(c^2_{sA}-w_A)\delta_A+3\mathcal{H}\left[3\mathcal{H}(1+w_A)(c^2_{sA}-w_A)+w'_A\right]\frac{\theta_A}{k^2}\nonumber\\
&+&(1+w_A)\theta_A+k^2(1+w_A)(B-E')-3(1+w_A)\psi'\nonumber\\
&=&a\frac{\bar{Q}_A}{\bar{\rho}_A}\left[\phi-\delta_A+3\mathcal{H}(c^2_{sA}-w_A)\frac{\theta_A}{k^2}\right]+a\frac{\delta Q_A}{\bar{\rho}_A},\\
\theta'_A&+&\mathcal{H}(1-3c^2_{sA})\theta_A-\frac{c^2_{sA}}{(1+w_A)}k^2\delta_A+\frac{2w_A}{3(1+w_A)}k^4\pi_A-k^2\phi\nonumber\\
&=&\frac{a\bar{Q}_A}{(1+w_A)\bar{\rho}_A}\left[\theta-(c^2_{sA}+1)\theta_A\right]-k^2f_A\frac{a}{(1+w_A)\bar{\rho}_A}.
\end{eqnarray}

In our decay vacuum case, one has the interaction and its corresponding perturbed term by comparing the background evolution equations of dark matter and dark energy
\begin{eqnarray}
\bar{Q}_c&=&-\bar{Q}_{de}=-\dot{\rho}_{de}=-3M^2_p a H \frac{d H}{d a}\sigma H_0,\\
\delta Q_c&=&-\delta Q_{de}=0,\\
f_c&=&-f_{de}=0.
\end{eqnarray}

Then the perturbed dark matter density contrast and velocity equations are given in longitudinal gauge as follows
\begin{eqnarray}
\delta'_c&=&-\theta_c+3\psi'-\frac{3a^3M^2_p}{\bar{\rho}_c}\frac{d H}{d a}\sigma H_0 \mathcal{H}(\phi-\delta_c),\\
\theta'_c&=&-\mathcal{H}\theta_c+k^2\phi.
\end{eqnarray}
For decay vacuum, it is not perturbed.

\section{Influence on CMB and matter power spectra} \label{sec:results}

\begin{figure}[!htbp]
\includegraphics[width=10cm]{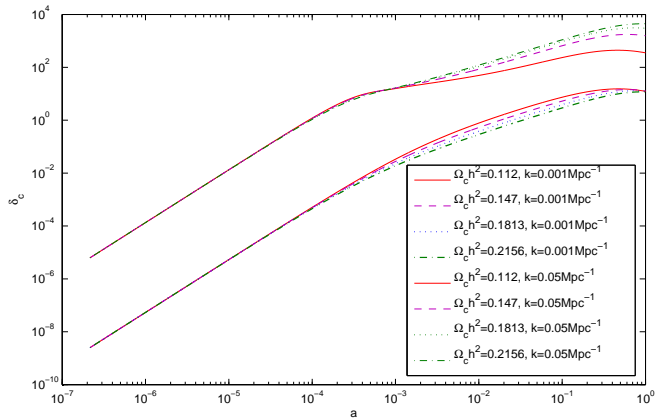}
\caption{The evolutions of cold dark matter density perturbation $\delta_{c}$ on scales $k=0.001\text{Mpc}^{-1}$, $k=0.05\text{Mpc}^{-1}$ with different values of $\omega_{c}$.}\label{fig:deltac}
\end{figure}

\begin{figure}[!htbp]
\includegraphics[width=8.8cm]{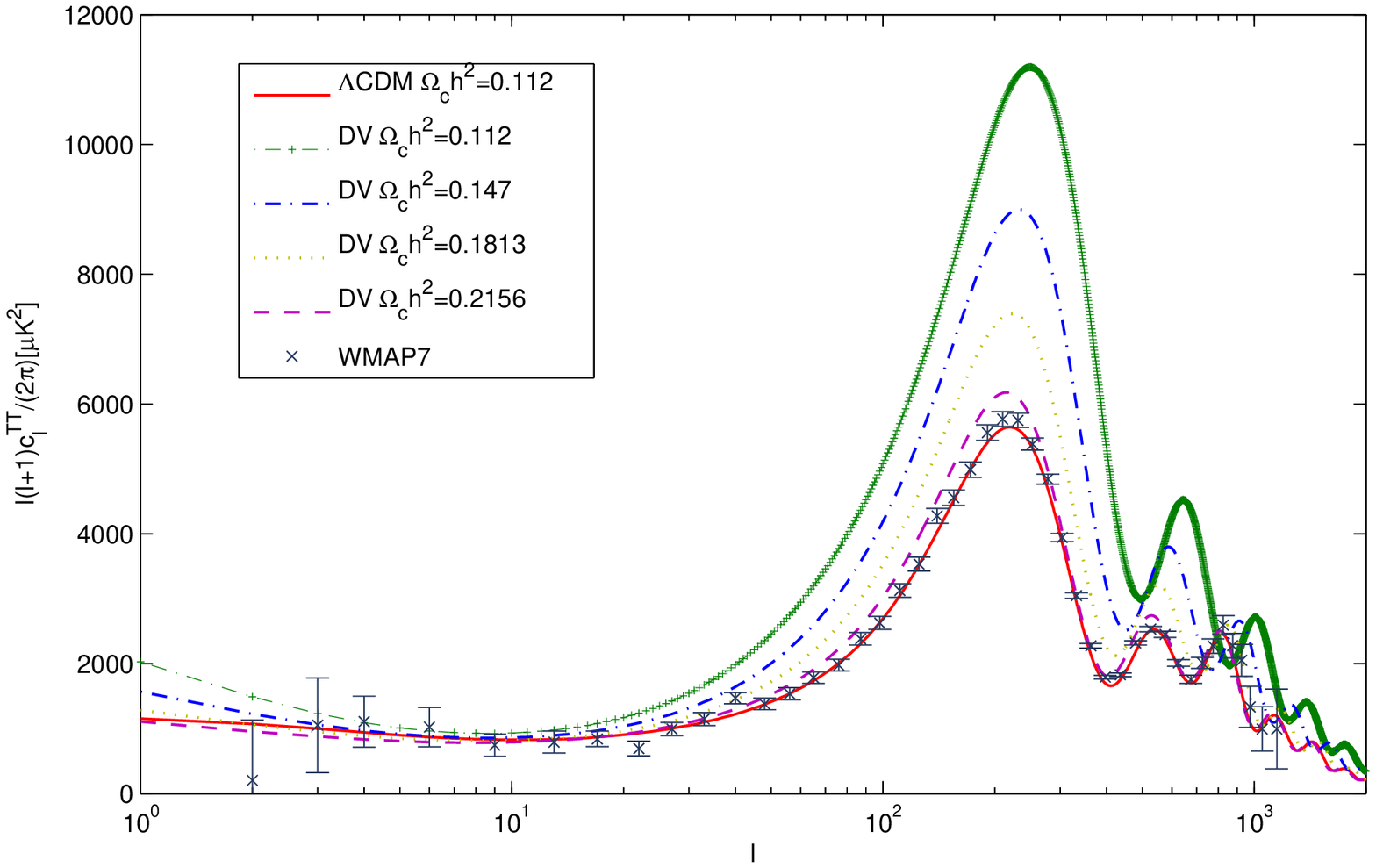}
\includegraphics[width=8.8cm]{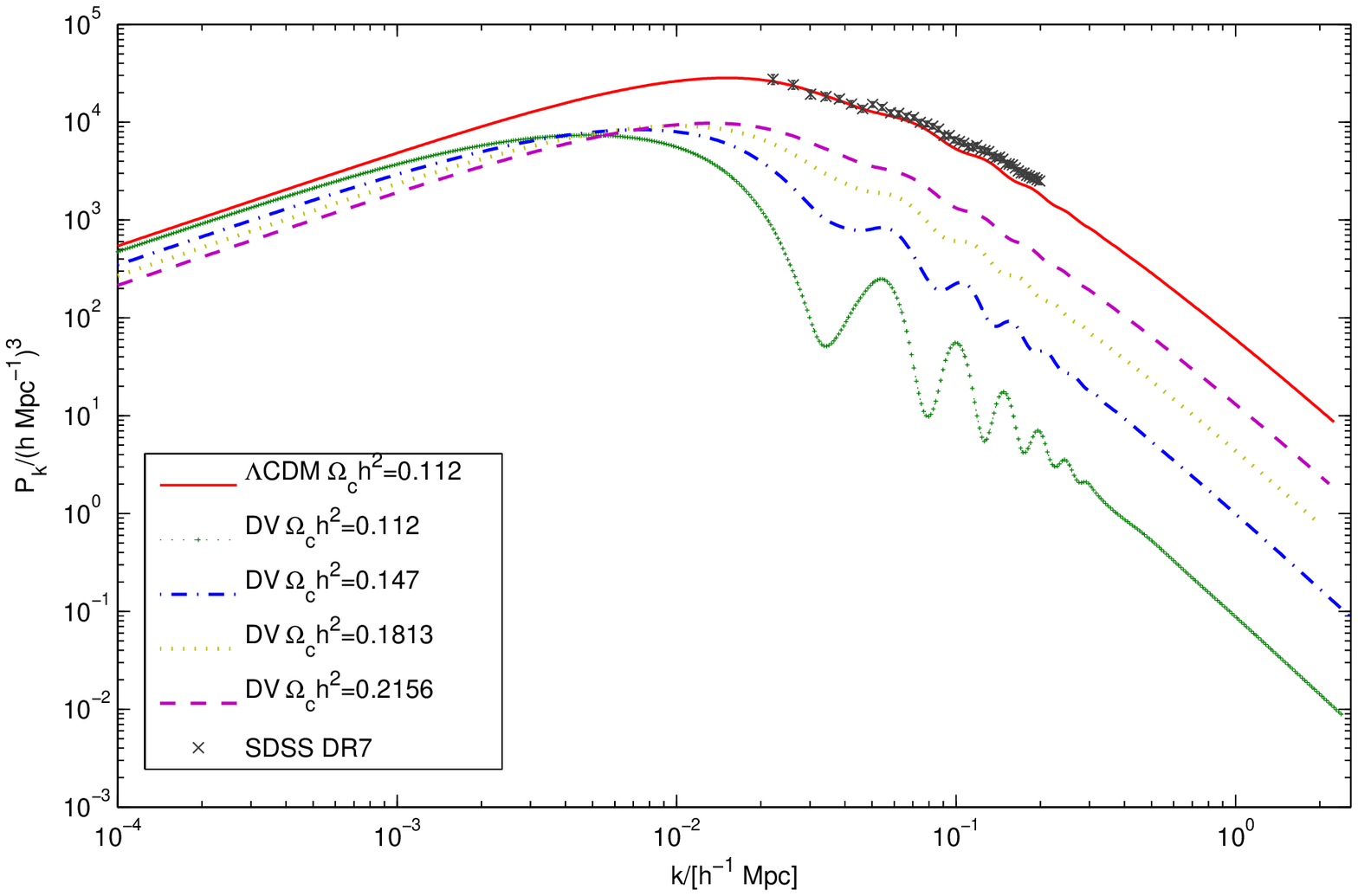}
\caption{The CMB temperature power spectrum and matter power spectrum with different values of $\omega_c$. The red solid lines correspond to $\Lambda$CDM model with $\omega_c=0.112$. The other lines correspond to time variable cosmological constant model with different values $\omega_c=0.112, 0.47, 0.1813, 0.2156$ respectively.}\label{fig:clspk}
\end{figure}

Now, we are in the position to study the CMB and matter power spectrum in this decay vacuum model. We modified the CAMB package \cite{ref:CAMB} to include the effective interaction between cold dark matter and time variable cosmological constant, and set to the adiabatic initial conditions. For comparison to $\Lambda$CDM model, we borrow the cosmological parameters values from WMAP7 \cite{ref:wmap7}. As outputs, the evolution of cold dark matter density contrast (Fig. \ref{fig:deltac}) on different scale $k=0.001, 0.05\text{Mpc}^{-1}$ and the CMB and matter power spectra (Fig. \ref{fig:clspk}) are shown for different values of $\Omega_{c}h^{2}=0.112, 0.147, 0.1813, 0.2156$, where $h=0.70$,
$\omega_b=0.0226$, $n_s=0.96$ and other relevant parameter values are fixed.

As shown in Fig \ref{fig:evolution}, decreasing the cold dark matter abundance $\omega_c$ will delay the matter-radiation equality time, the resultant acoustic peak will be enhanced Fig. \ref{fig:clspk}. In our case, the acoustic peak is enhanced also due to the interaction between dark sectors. 
From Fig. \ref{fig:clspk}, one can see that the CMB temperature spectrum favors a large abundance of cold dark matter. At low $l$ part where the integrated Sachs-Wolfs effect is dominated, one can read from figure that the ISW effect can not distinguish the decay vacuum model from $\Lambda$CDM model. As expected, the matter (baryon) power spectra are really different from $\Lambda$CDM model because of its large abundance of $\Omega_{b}$ during the evolution as shown in the right panel of Fig. \ref{fig:evolution}. With these observations, one can conclude that CMB observations and matter power spectrum can distinguish the decay vacuum model from $\Lambda$CDM model remarkably.

\section{A generalized form $\bar{\rho}_{\Lambda}=3M_p^2(\xi_1 H_0 H+\xi_2  H^2)$}\label{sec:tvv}

A generalized form, dubbed as a time variable cosmological constant or vacuum energy, is assumed \cite{ref:ISW3,ref:TVV}
\begin{equation}
\bar{\rho}_{\Lambda}=3M_p^2(\xi_1 H_0 H+\xi_2  H^2),
\end{equation}
Here an extra $H_0$ is introduced in the first coupling term for convenience. Then, $\xi_1$ and $\xi_2$ are dimensionless parameters. $\xi_1=0$ case corresponds to Holographic dark energy model with IR cut-off $H^{2}$ \cite{ref:holo}. And when $\xi_2=0$, it reduces to decay vacuum model \cite{Borges1}.  Following the calculations in section \ref{sec:bg}, one has the relation
\begin{equation}
\xi_1+\xi_2=1-\Omega_{r0}-\Omega_{b0}-\Omega_{c0}.
\end{equation}
and the fraction of dark energy
\begin{equation}
f_{de}=\frac{\bar{\rho}_{de}}{\bar{\rho}_{de0}}\equiv\frac{\xi_1 H/H_0+\xi_2 H^2/H^2_0}{\xi_1+\xi_2}.
\end{equation}
Substituting the above relation of $f_{de}$ into Eq. (\ref{eq:FRE}), one has a quadratic equation of $H/H_0$. After a simple algebra, one obtains
\begin{equation}
H=\frac{1}{2\alpha}H_0(\xi_1\Omega_{de0}+\gamma)\label{eq:H2},
\end{equation}
where
\begin{eqnarray}
\alpha&=&\xi_1+\xi_2(\Omega_{r0}+\Omega_{b0}+\Omega_{c0})\\
\gamma&=&\sqrt{\xi^2_1\Omega^2_{de0}+4\alpha(\xi_1+\xi_2)(\Omega_{r0}a^{-4}+\Omega_{b0}a^{-3}+\Omega_{c0}f_c(a))}.
\end{eqnarray}
Here, the negative solution is removed. Solving quadratic equation of $H/H_0$ and keeping the positivity of root, the constraint condition $\xi_{2}\le 2-\Omega_{de0}$ or $\xi_{1}+2\xi_{2}\le 2$ is respected. Also, the positivity of Eq. (\ref{eq:H2}) requires $\alpha>0$, i.e $\xi_{2}<1$. At last, one has $0<\xi_{1}+\xi_{2}<1$ and $\xi_{2}<1$. The function $f_c(a)$ is a solution of the differential equation
\begin{equation}
\frac{d f_c}{d\ln
a}=\frac{[\xi_1\alpha\Omega_{de0}+\xi_2(\xi_1\Omega_{de0}+\gamma)\Omega_{de0}](3\Omega_{b0}a^{-3}+4\Omega_{r0}a^{-4})-3\alpha\gamma\Omega_{c0}f_c}{\alpha\gamma\Omega_{c0}+\xi_1\alpha\Omega_{c0}\Omega_{de0}+\xi_2\Omega_{c0}\Omega_{de0}(\xi_1\Omega_{de0}+\gamma)}
\end{equation}
with current value $f_c(1)\equiv1$. It is easy to obtain the conventional dark matter evolution equation $f_c(a)=a^{-3}$, when the cosmological constant is a real constant. 

At very early epoch (for example $a\sim 10^{-6}$), the differential equation of $f_{c}$ reduces to
\begin{equation}
\frac{d f_c}{d\ln a}\approx \frac{4\xi_{2}\Omega_{r0}a^{-4}-3\Omega_{c0}(1-\xi_{2})f_c(a)}{\Omega_{c0}}
\end{equation}
which has solution
\begin{equation}
f_c(a)\approx a^{3(-1+\xi_2)} C-\frac{4 \xi _2 \Omega _{r0}a^{-4}}{\left(1+3 \xi _2\right) \Omega _{c0}},
\end{equation}
where $C>0$ is an integration constant. To keep the positivity of $f_{c}(a)$, the stringent constraint $-1/3< \xi_{2} \le 0$ is respected. However, when $\xi_{2}$ is a negative dimensionless parameter, the second term of $\rho_{\Lambda}$ will be dominated in the early universe. And a negative vacuum energy or dark energy density will appear. That is prohibited. Actually, in this case, to keep the positivity of vacuum energy density $\rho_{\Lambda}>0$, one has 
\begin{equation}
\frac{\Omega _{de0}}{1-H/H_{0}}<\xi _2\le 0.
\end{equation}
So, the parameter space of $\xi_{2}$ depends on the Hubble parameter values in the early epoch. Then, a fine tuning problem would be committed. It means that the value of $\xi_{2}$ is a very small negative parameter, i.e. $\xi_{2}\sim 0$. In this sense, it would not be a viable dark energy model with the exception of $\xi_{2}=0$. Of course, one may argue that in that early epoch the assumption of this time variable cosmological constant model is blown up. The above analysis is based on a basic physical reality that is the positivity of energy density $\rho_{i}>0$ in the whole evolution of the universe. If one can accept the fine tuning, the parameter space of $\xi_{2}$ would be in the range $[-\epsilon, \epsilon]$ where $\epsilon$ is very small dimensionless constant to keep the positivity of energy densities of $\rho_{c}$ and $\rho_{\Lambda}$, for example $\epsilon \sim 10^{-6}$. So, to avoid the 'unnatural' fine tuning, $\xi_{2}$ would be zero. Then the time variable cosmological constant model reduces to the so-called decay vacuum model
\begin{equation}
\bar{\rho}_{\Lambda}=3M_p^2\xi_1 H_0 H. 
\end{equation} 
In this sense, this generalized form is not a viable dark energy model.

\section{Conclusion}\label{sec:con}

In this paper, a decay vacuum model $\bar{\rho}_\Lambda=3\sigma M_p^2H_0 H$ and its generalization $\bar{\rho}_\Lambda=3M_p^2(\xi_1 H_0 H+\xi_2  H^2)$, a time variable cosmological constant model, are revisited.  At first, the background evolution equation in a spatially flat FRW universe containing cold dark matter, radiation, baryon and time variable cosmological constant is given. The relative departure from $\Lambda$CDM model is minor, please see the left panel of Fig. \ref{fig:evolution}. So to discriminate the decay vacuum model from $\Lambda$CDM model, high redshift observations are needed. In the decay vacuum model case, an effective interaction between cold dark matter and vacuum can be introduced. Then the evolution of cold dark matter will depart from the conventional power law $a^{-3}$. And the large scale structure formation will be strongly different from that of $\Lambda$CDM model. Though the baryon component evolves in the scaling $a^{-3}$, the background evolves different from $\Lambda$CDM model for the effective interaction between cold dark matter and decay vacuum. Then the dynamic evolution would be modified. So the cosmological perturbations are taken into account. As results, the angular power spectrum of CMB and matter power are presented with different parameter values of cold dark matter abundance $\omega_c$, please see Fig. \ref{fig:clspk}. From this figure, one can conclude that CMB observations and matter power spectrum can distinguish the decay vacuum model from $\Lambda$CDM model markedly. When $\Omega_{c} h^{2}=0.2158$, i.e. $\Omega_{c0}=0.4404$, the purple dashed line in Fig. \ref{fig:clspk} is close to observational data points. It means that increasing the abundance of cold dark matter will depress the acoustic peaks to cosmic observational data points in this model. However, in the right panel of Fig. \ref{fig:clspk}, one sees that increasing the abundance of cold dark matter will enhance the matter power spectrum at small scale but depress that at large scale. That makes it difficult to match observational data points. With these observations, this model would be ruled out. But to know in what kind of levels to rule out this model, testing this model with current available cosmic observational data sets, for example type Ia supernovae, baryon acoustic oscillation, full CMB and SDSS DR7 etc, would be interesting. Furthermore, a generalized vacuum model $\bar{\rho}_{\Lambda}=3M_p^2(\xi_1 H_0 H+\xi_2  H^2)$ was discussed. From a detailed analysis, one can find that the parameter space of $\xi_{2}$ is a very small negative dimensionless parameter. To keep the positivity of energy density of dark matter and dark energy at early epoch, the parameter $\xi_{2}$ suffers from the fine tuning problem. So to avoid this unnatural condition, the $\xi_{2}$ would be set to zero. Then it reduces to the decay vacuum model. In this sense, it would not be a viable dark energy model.

\acknowledgements{We thank Prof. Jai-chan Hwang and Dr. Chan-Gyung Park for useful discussion and anonymous referee's invaluable help to improve the manscript. L. Xu's work is supported by NSF (10703001) of P. R. China and the Fundamental Research Funds for the Central Universities (DUT10LK31). H. Noh's work is supported by Mid-career Research Program through National
Research Foundation funded by the MEST (No. 2010-0000302).}

\end{document}